\documentclass[aps,prl,twocolumn,superscriptaddress]{revtex4-1}

\usepackage{graphicx}

\usepackage{amssymb,amsfonts,amsmath}

\begin{document}





\title{The morphodynamics of 3D migrating cancer cells}

\author{Christopher Z. Eddy}
\affiliation{Department of Physics, Oregon State University, Corvallis OR, 97331}
\author{Xinyao Wang}
\affiliation{School of Electrical Engineering and Computer
	Science, Oregon State
	University, Corvallis OR, 97331}
\author{Fuxin Li}
\affiliation{School of Electrical Engineering and Computer
	Science, Oregon State
	University, Corvallis OR, 97331}
\author{Bo Sun}
\affiliation{Department of Physics, Oregon State University, Corvallis OR, 97331}


\begin{abstract}
Cell shape is an important biomarker that is directly linked to cell
function. However, cell morphodynamics, namely the temporal fluctuation
of cell shape is much less understood.  We study the morphodynamics of
MDA-MB-231 cells in type I collagen extracellular matrix (ECM).  We
find ECM mechanics, as tuned by collagen concentration, controls the
morphodynamics but not the static cell morphology. We employ machine learning to classify cell shape into five
different morphological phenotypes corresponding to different migration modes. As a result, cell morphodynamics is mapped into temporal evolution of morphological phenotypes. We systematically
characterize the phenotype evolutions including occurrence probability,
dwell time, transition flux, and 3D migrational characteristics. We find that manipulating Rho-signaling enhances the morphodynamics and phenotype transitions. Using a tumor organoid model, we
show that the distinct invasion potentials of each phenotype modulate
the phenotype homeostasis. Overall invasion of a tumor organoid is
facilitated by individual cells searching for and committing to phenotypes of
higher invasive potential. In conclusion, we show that 3D migrating
cancer cells exhibit rich morphodynamics that is regulated by ECM
mechanics, Rho-signaling, and is closely related with cell motility. Our results pave
the way to the systematic characterization and functional understanding of
cell morphodynamics as a new biomarker for normal and malignant cells.
\end{abstract}


\maketitle

\section{Introduction}
\label{sec:Intro}
Shape defines the cell. In the 1677 book \textit{Micrographia}, Robert
Hooke showed sections within a herbaceous plant under a
microscope. The shape of those sections resembles cells in a
monastery, so he named the structures cells \cite{Hooke1665}. Many
breakthroughs followed Hooke's discovery, from the cell theory of
Schwann and Schleiden, to the theory of tissue formation by Remak,
Virchow and Kolliker, and the theory of cellularpathologie by Virchow,
all of which are inspired by observations of cell shapes, or
morphology in general \cite{Mazzarello1999,Mayr1982}.

In our modern view cell shape is determined by cell function
\cite{Walter2014,Ingber1994}. A nerve cell has long branched
protrusions for communication with other neurons; while the cuboidal shape
of epithelial cells allow them to tile the surface of organs. Loss of
characteristic shape, on the other hand, is associated with functional
abnormality. Thus morphological characterization has been an important
tool for diagnosis such as in red blood cell disease \cite{Diez2010},
neurological disease \cite{Serrano2011}, and cancer
\cite{Bakal2013,Wu2015}. More recently, cell shape analysis is boosted
by techniques from computer vision. As a result, it becomes possible
to obtain high content information of cellular states from
morphological data alone \cite{Perrimon2007,Wu2015,Lam2017,Carpenter2017}.

While most research focuses on the static cell morphology, the dynamic
fluctuation of cell shape is much less understood. However, shape
fluctuation -- namely morphodynamics, is of central importance for
dynamic cellular functions. The abnormal diffusion of small
protrusions - microvilli - on the surface of a T cell allows the T
cell to efficiently scan antigen-presenting surfaces
\cite{Caieaal3118}. For a migrating cancer cell, morphodynamics drives
the motility of the cell in many ways similar to our body frame
movements that enable swimming. In fact, just as there are different
swimming styles, cancer cells have been observed to execute six
different programs during invasion in 3D tissue space
\cite{Konstantopoulos2017}. Each program has distinct signatures of
morphology and morphodynamics, and are usually referred to as
migration phenotypes of filopodial, lamellipodial, lobopodial,
hemispherical blebbing, small blebbing, and actin-enriched leading
edge \cite{Petrie2012}. Cancer cell migration phenotypes are controlled
by intracellular signaling such as the Rac and Rho-Rock-myosin pathways
\cite{Marshall2008rac,Marshall2005rho}, and extracellular factors such
as the elasticity, and degradability of the extracellular matrix (ECM)
\cite{Marshall2010plasticity,Yamada2012commentary}. The ability of a
cancer cell to switch between migration phenotypes is important for
tumor prognosis. Many therapies, such as MMP inhibitors that target a
particular mode of cell motility, fail to stop tumor metastasis largely
because cells take other available migration programs
\cite{Zucker2003,Friedl2003MMP_inhibition_transition}.

In this paper, we study the morphodynamics of MDA-MB-231 cells, a
highly invasive human breast cancer cell line in 3D collagen
matrices. We find the shape fluctuation is regulated by the mechanics
of ECM. The cellular morphodynamics not only drives 3D migration, but
also allows a single cell to sample multiple morphological phenotypes
over a short amount of time. As a result, ECM mechanics regulate the
stability and transitions between morphological phenotypes that
corresponding to different migration programs. We have measured the
motility of each morphological phenotype and find that phenotype transition
facilitates invasiveness during 3D tumor organoid invasion.

\begin{figure}[h]
	\centering \includegraphics[width=0.99\columnwidth]{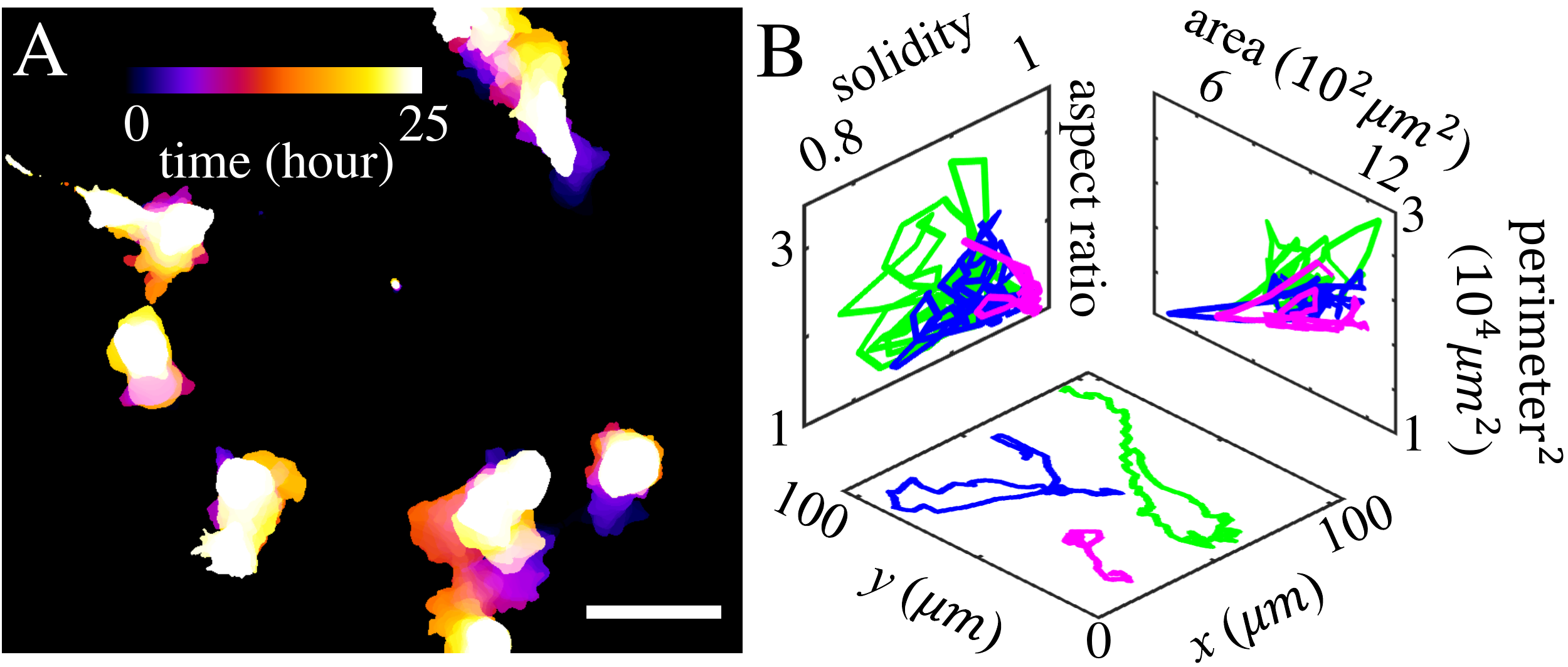}
	\caption{Three-dimensional migration of MDA-MB-231 cells is
		accompanied with significant cell shape fluctuation. (A) A typical
		time lapse recording of 25 hours is projected onto a single image with colors
		representing time. (B) The real space (x-y plane), and shape space
		trajectories of 3 cells shown in (A).}
	\label{fig1}
\end{figure}
\section{Results}
\label{sec:results}

We find 3D migrating cancer cells rapidly sample their available shape
space. In order to quantify the cell morphodynamics, we take
time-lapse fluorescent images of MDA-MB-231 cells migrating in
collagen matrices. The GFP labeled cells typically stay within the
focal depth of the objective lens (20X, NA 0.7) for 10-20 hours, while
we obtain 2D cell images at a rate of 4 frames per hour (see \textit{SI Appendix} section S1). After
binarization and segmentation, we compute geometric measures such as area,
perimeter, aspect ratio, and solidity which collectively quantify the
shape of a cell (see \textit{SI Appendix} section S2).

\begin{figure}[h]
	\centering \includegraphics[width=0.99\columnwidth]{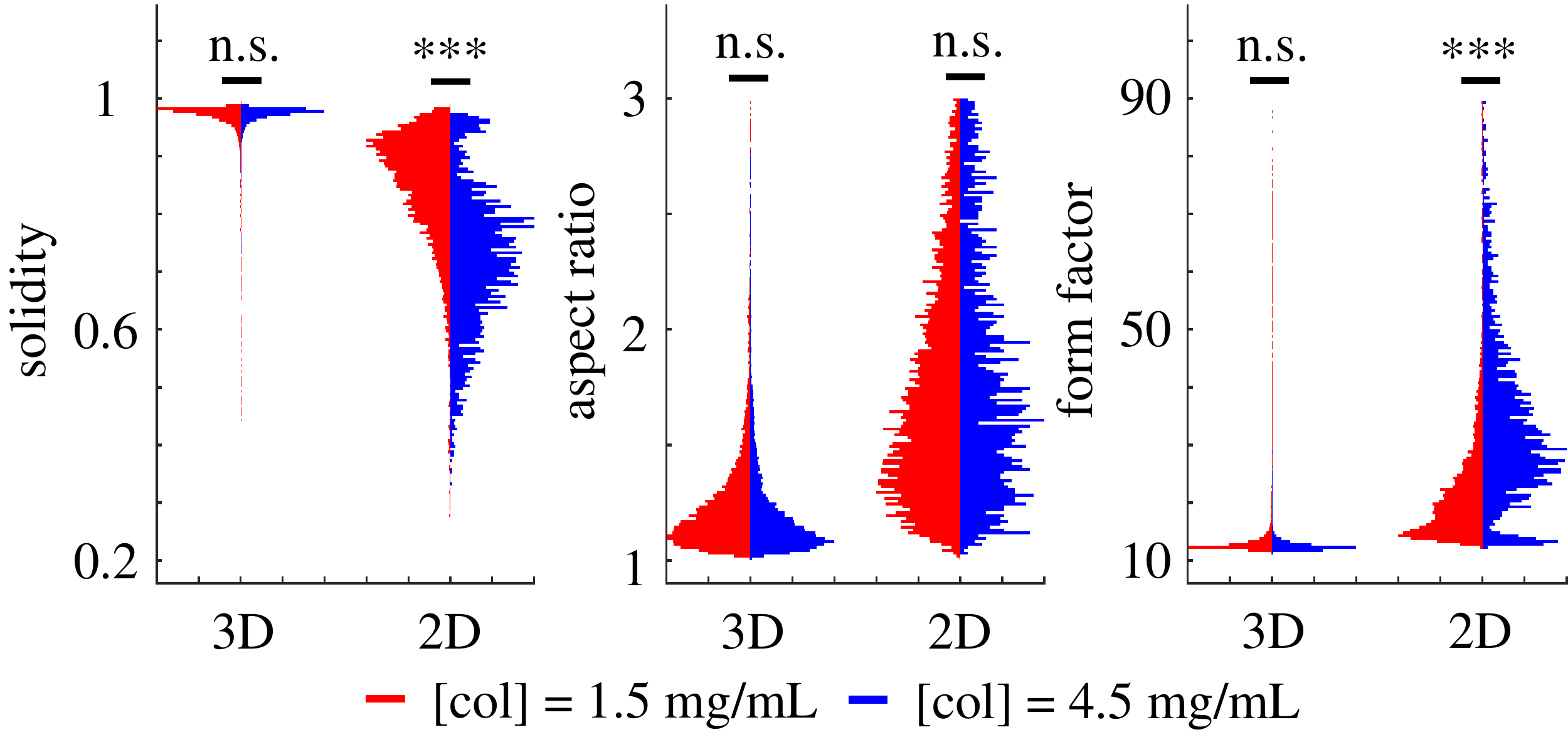}
	\caption{Ensemble distribution of shape characteristics of MDA-MB-231
		cells in 3D (embedded in collagen gels) and 2D (on the top surfaces
		of collagen gels) cultures. The histograms show relative
		probabilities (peak normalized to one) of solidity, aspect ratio,
		and form factor. For each distribution sample size $N>$ 2000. ***:
		$p<$0.001, n.s.: not significant (ANOVA). Color of the bars
		represent collagen concentration, red: 1.5 mg/mL, and blue: 4.5
		mg/mL.}
	\label{fig2}
\end{figure}
The morphodynamics of a cell manifests itself as a random walk in the
geometric shape space concurrent with its motility in the 3D matrix
(Fig. \ref{fig1}). Because any cell shape change boils down to the
physical interactions between the cell and its environment, we hypothesize that morphodynamics is sensitive to the physical
properties of the ECM. To test the hypothesis, we examine the shape
fluctuation of thousands of MDA-MB-231 cells in type I collagen
matrices of varying concentrations (see \textit{SI Appendix} section S2). As we increase collagen
concentration from 1.5 mg/mL to 4.5 mg/mL, the storage modulus
increases by more than three folds (see \textit{SI Appendix} section S3).

We first examine if ensemble distributions of cell shape shift as
collagen concentration increases. To this end we focus on three
quantifies: solidity (ratio of the area and its convex area), aspect
ratio (ratio of the major and minor axes lengths), and form factor
(ratio of squared perimeter and area) that are independent of imaging
scales. As shown in Fig. \ref{fig2}, these quantities are sharply
distributed for 3D migrating cells. However, none of the three
quantities show significant changes when collagen concentration
increases by 3-fold. This is in clear contrast to MDA-MB-231 cells
migrating on 2D surfaces. Consistent with previous reports, when
collagen concentration increases from 1.5 mg/mL to 4.5 mg/mL,
MDA-MB-231 cells exhibit lower solidity and higher form factor, as a
result of enhanced integrin-mediated adhesion and membrane protrusions
\cite{Janmey2005}.

\begin{figure}[h]
	\centering \includegraphics[width=0.99\columnwidth]{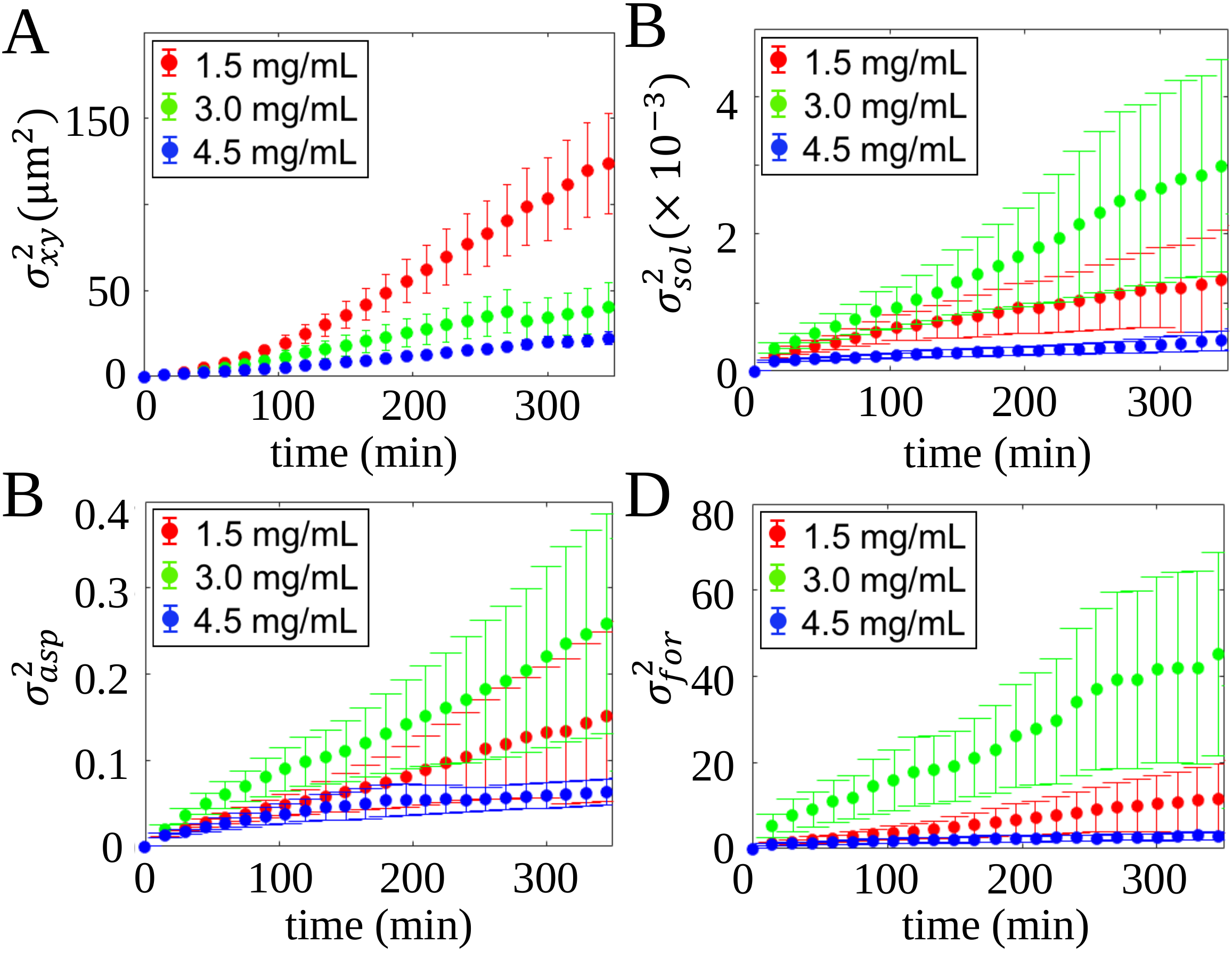}
	\caption{The mean square displacement (MSD) $\sigma^2$ of MDA-MB-231 cells in real space and
		in shape space as a function of time $t$. (A) MSD of real space
		motility. $\sigma_{xy}^2$ is
		obtained from the x-y components of 3D cell trajectories. (B-D) MSDs
		characterize the fluctuation of solidity ($\sigma_{sol}^2$), aspect
		ratio ($\sigma_{asp}^2$), and form factor ($\sigma_{for}^2$). For
		each collagen concentration about 200 cells are analyzed. }
	\label{fig3}
\end{figure}

While the ensemble distributions of cell shape do not distinguish between
different collagen concentrations in 3D cultures, we find
morphodynamics do. For 3D migrating cells we calculate their mean
square displacements in real space ($\sigma_{xy}^2$ in
Fig. \ref{fig3}A) and in shape space ($\sigma_{sol}^2$,
$\sigma_{asp}^2$, and $\sigma_{for}^2$ in
Fig. \ref{fig3}B-D). MDA-MB-231 cells show higher diffusivity in real
space at lower collagen concentration, as is expected when matrix pore
size is rate-limiting for cell migration \cite{Friedl2013physicallimits,Sapudom2015}.

Remarkably, cellular morphodynamics show clear dependence on collagen
concentrations. Over the time scale of a few hours, $\sigma_{sol}^2$,
$\sigma_{asp}^2$, and $\sigma_{for}^2$ are approximately linear
functions of time. Interestingly, diffusivity of solidity, aspect
ratio, and form factor are highest at intermediate collagen
concentration (3.0 mg/mL), and are strongly suppressed at collagen
concentration of 4.5 mg/mL.

The fact that ECM mechanics significantly regulates cellular
morphodynamics, but not static cell shapes, may be understood from a
simple physical picture. The morphodynamics of a cell may be
considered as a random walk over a complex potential landscape
$\mathcal{U}_{shape}$ defined over the cell shape space. At
equilibrium, the distribution of static cell shape (Fig. \ref{fig2})
is determined by $\mathcal{U}_{shape}$ while the speed of fluctuation
(Fig. \ref{fig3}) is also influenced by the effective
viscosity. Keeping this simple picture in mind, we conclude that for
2D migration ECM concentration regulates the topography of the
potential landscape while in 3D ECM concentration mainly affects
the effective viscocity.

Motivated by the mechanosensitivity of cell shape fluctuation, we next seek to gain more
insights by investigating a course-grained version of
morphodynamics. To this end, we train machine classifiers to divide
the high dimensional cell shape space based on 3D cell migration
phenotypes. This is possible because different migration modes are
associated with distinct characteristic cell morphologies \cite{Yamada2012commentary,Konstantopoulos2017}. 

We consider five morphological phenotypes including two mesenchymal
ones: filopodial (or FP in short) and lamellipodial (LP); as well as
three amoeboidal ones: hemispherical blebbing (HS), small blebbing
(SB), and actin-enriched leading edge (AE). A sixth phenotype, namely
lobopodial or nuclear piston mode, has not been observed in our
experiments which is consistent with previous reports \cite{Yamada2017restore}. Once
the classifier is trained, phenotype is determined automatically from
a cell image if a particular phenotype receives more than 70\%
probability score. On the other hand, if none of the five phenotypes
receive more than 70\% probability score, the cell is considered to be
in an intermediate state.

\begin{figure}[h]
	\centering \includegraphics[width=0.99\columnwidth]{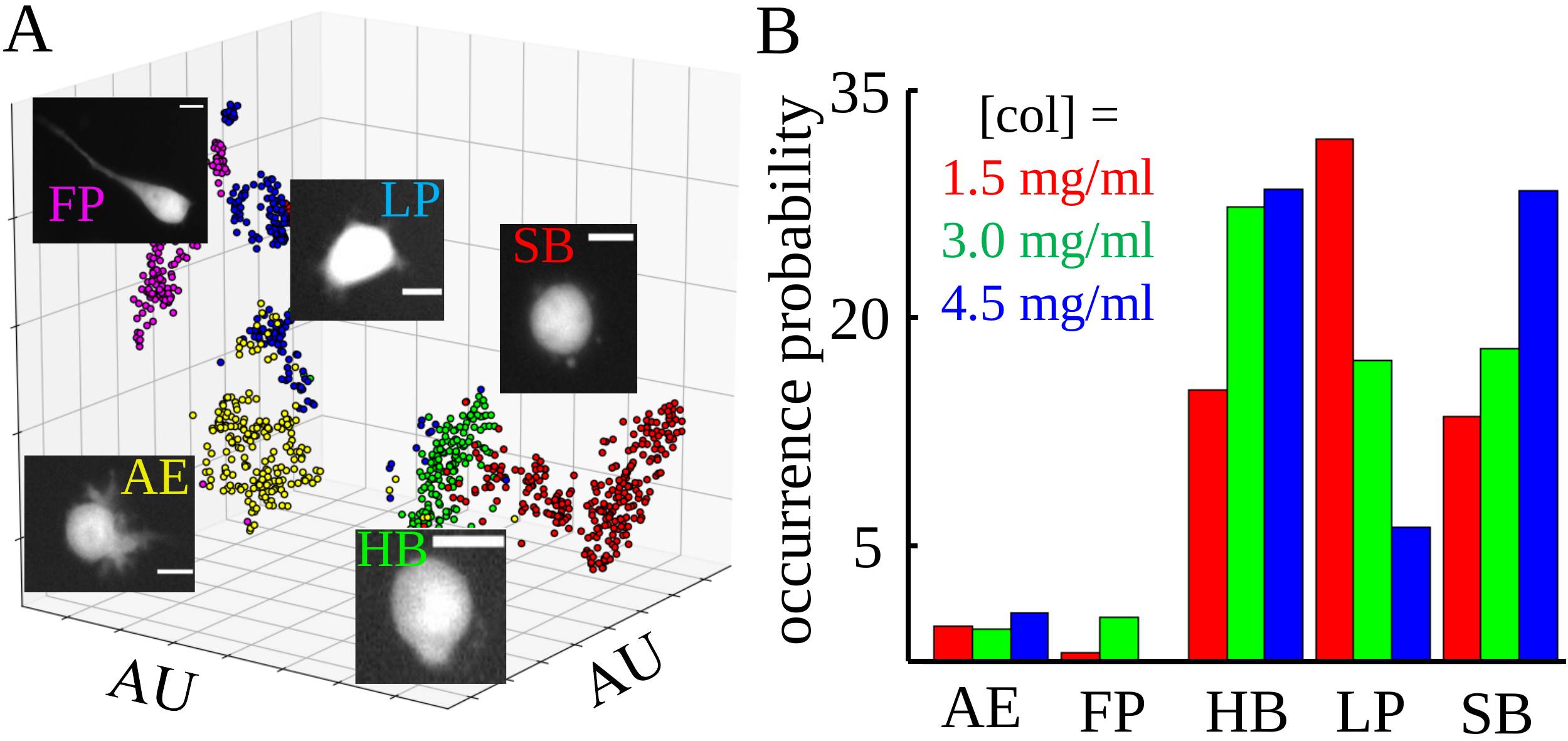}
	\caption{MDA-MB-231 cells exhibit multiple morphological phenotypes
		that are typically associated with distinct 3D migration modes. (A)
		An SVM classifier is trained based on geometry of binarized cell
		images. A threshold of 70\% probability is required for
		each phenotype classification, or the cell is classified to be in
		intermediate state. To better visualize the high dimensional geometric measures, we apply t-SNE method \cite{Maaten2009} to generate a 3D projection of the cell shape spaces. Different morphological phenotypes (insets show representative images) are clearly separated. Scale bars of the insets are 20 $\mu$m. (B)
		Percentage occurrence of each morphological phenotype (intermediate
		states account for approximately 30\% occurrence in each collagen concentration). AE: actin-enriched leading edge. FP:
		filopodial, LP: lamellipodial, HS: hemispherical blebbing, SB:
		small blebbing. Colors represent collagen concentration, red: 1.5
		mg/mL, green: 3 mg/mL, blue: 4.5 mg/mL. For each collagen
		concentration more than 3000 single cell images are analyzed.}
	\label{fig4}
\end{figure}

We have trained two classifiers (see \textit{SI Appendix} S4). The first one is based on support vector machines (SVM
\cite{Cortes1995,Ben-Hur2001}) involving 66 geometric measures. The
second one is based on a convolutional neural network architecture
that uses raw gray scale images as inputs \cite{Ben-Hur2016}. Both
classifiers show good separation of training sets, more than 90\% of
successful rates, and agree with each other well on test data sets
(Fig. \ref{fig4}A and \textit{SI Appendix} S4). Following, we
mainly report the results from SVM algorithm since it has intuitive
interpretations directly from the geometric measures.

The likelihoods of observing different morphological phenotypes are
remarkably different. At $[col]$ = 1.5 mg/mL, HB, LP and SB shapes
each account for more than 15\% of the observations, FP and AE
phenotypes only occur less than 5\% of the time. We find increasing
collagen concentration increases the occurrence of HS and SB cells,
while decreasing the occurrence of LP cells. This observation suggests
that phenotype homeostasis of MDA-MB-231 cells is regulated by ECM
mechanics such that stiffer ECM promotes higher probability of
amoeboidal phenotypes.

\begin{figure}[h]
	\centering \includegraphics[width=0.99\columnwidth]{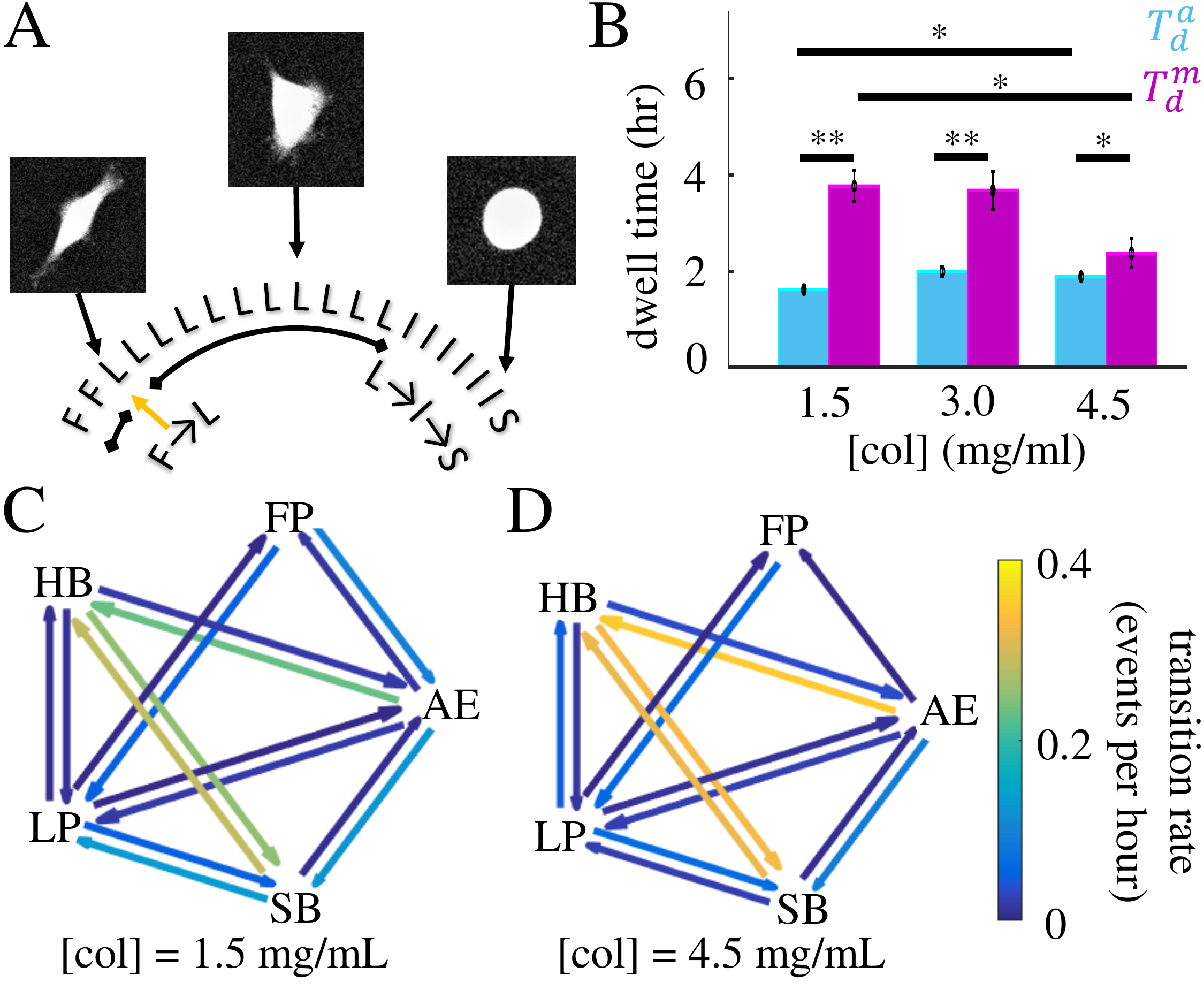}
	\caption{Collagen concentration regulates the morphological phenotype
		homeostasis of 3D migrating MDA-MB-231 cells. (A) A sample time
		series of morphological phenotype. Insets: three snapshots showing
		the GFP-labeled cell. Abbreviations: F -- filopodial, L --
		lamellipodial, I -- intermediate state. (B) The dwell times of
		amoeboidal ($T_d^a$, cyan) and mesenchymal ($T_d^m$, magenta) cell
		shapes have opposite dependence on collagen concentration. (C-D)
		Graphical representation of the transition matrix between different
		morphological phenotypes at collagen concentration of 1.5 mg/mL (c)
		and 4.5 mg/ml (D). The results of (B-D) are obtained by analyzing a
		total of more than 6200 hours (four frames per hour) of single cell
		morphodynamic trajectories. See also \textit{SI Appendix} S5 for
		additional details.}
	\label{fig5}
\end{figure}

To obtain further insights of cell morphodynamics, we investigate the
time evolution of morphological phenotypes. Fig. \ref{fig5}A shows a
typical time series of phenotype dynamics where a cell switch directly
from FP to LP shape, then to SB shape via intermediate state. We first
examine the overall stability of cell shapes by measuring the average
dwell time a cell staying continuously either in amoeboidal or
mesenchymal states (see also \textit{SI Appendix} S5). As shown in
Fig. \ref{fig5}B, amoeboidal dwell time $T_d^a$ slightly increases
with collagen concentration, while mesenchymal dwell time $T_d^m$
decreases. At concentration $[col]$ = 1.5 mg/mL, $T_d^m$ = 3.8 hrs is
more than twice longer than $T_d^a$ = 1.6 hrs. When collagen concentration
increases to 4.5 mg/mL, mesenchymal shapes become less stable, and the
corresponding dwell time (2.4 hrs) is only 30\% longer than the amoeboidal dwell time (1.8 hrs).

To reveal the details of phenotype dynamics, we have computed the
transition rates as shown in Fig. \ref{fig5}C-D. Surprisingly, we
notice that while higher ECM concentration extends the amoeboidal
dwell time, it doesn't slow down the phenotype dynamics at
all. Instead, we observe more frequent transitions along the HB-SB-AE
amoeboidal axes. For instance, the transition rates between HB and SB
states at $[col]$ = 4.5 mg/mL is 20\% higher than the rates at $[col]$
= 1.5 mg/mL. While on the contrary, the transition rates along the
mesenchymal axis FP-LP barely change when collagen concentration
increases by 3 folds.

Results in Fig. \ref{fig5}C-D also reveal that the reduction of
mesenchymal dwell time $T_d^m$ at higher collagen concentration is
mainly due to the two-fold increase of transition rates from
lamellipodial to amoeboidal states. This is consistent with the
mechanical model of blebbing formation
\cite{Tinevez2009,Yamada2012nonpoloarized}. A lamellipodial cell
exhibits a prominent cortical F-actin layer. Blebbs form when
actomyosin contractility exceeds the binding between cortical actin
and cell membrane. Our results suggest that the two competing forces
exhibit higher fluctuation at increased ECM concentration, fueling the
transition between lamellipodial and blebbing states.

On the other hand, we observe that filopodial cells have very low
transition rates to and from blebbing phenotypes. Filopodial
protrusions consist of elongated and bundled actin fibers as a result
of elevated actin polymerization and cross-linking by Ena/VASP
proteins \cite{Borisy2004}. Fig. \ref{fig5}C-D suggest that the mechanical
barrier separating filopodia and blebbing protrusions is too high
for the actomyosin contractility to overcome directly. Instead, a FP
cell can turn into a blebbing shape by first transform into AE or LP
states.
\begin{figure}[h]
	\centering \includegraphics[width=0.99\columnwidth]{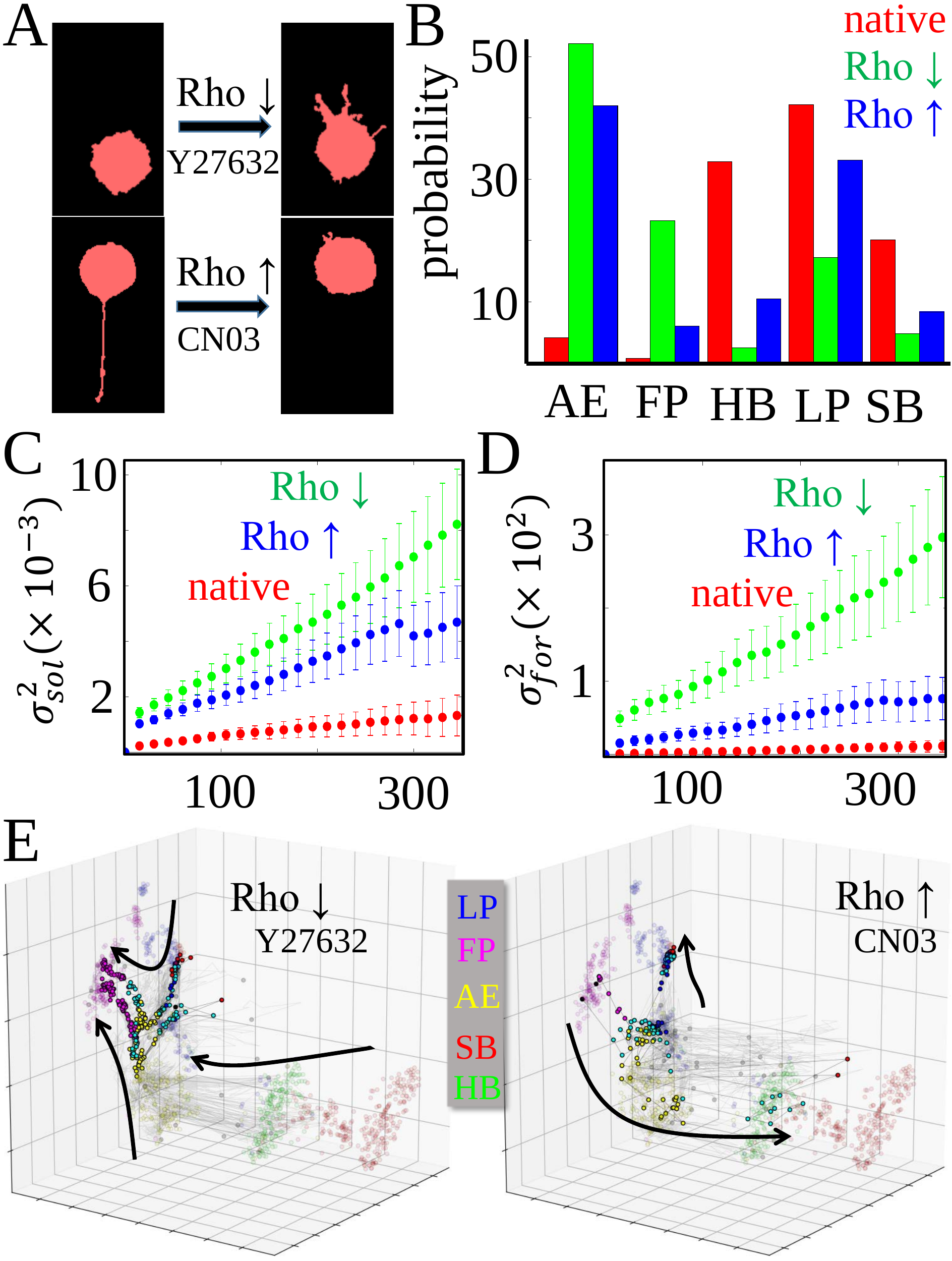}
	\caption{Morphodynamics of 3D cultured MDA-MB-231 cells under pharmacological perturbations of Rho-signaling. (A) Representative morphological changes under treatment of Y27632 or CN03. (B) The occurrence probabilities of each morphological phenotypes after 8 hours of Y27632 (Rho$\downarrow$) and CN03 (Rho$\uparrow$) treatments. (C-D) The mean square displacement $\sigma^2$ of solidity (C) and form factor (D) for native (non-treated), Y27632-treated, and CN03-treated cells. (E) Trajectories (thin gray lines) of cells in the t-SNE embedded shape space. The trajectories start immediately after introducing Y27632 or CN03, and ends after 8 hours of incubating with the drugs. Converging paths are indicated with thick curves with arrows as guide to the eyes. Two representative trajectories per each treatment are highlighted with colored dots connected by black lines, where color represents the instantaneous phenotype. Scattered light-colored dots show training sets which is the same as in Fig. \ref{fig4}A. For each treatment approximately 100 cells are analyzed for their morphodynamics.} 
	\label{fig6}
\end{figure}

In order to understand the mechanisms underlying cell morphological phenotype transitions, we examine the effects of manipulating Rho-signaling, which is a master regulator that determines the mechanical state of a cell. To this end, we apply Y27632, a Rho-inhibitor; and CN03, a Rho-activator to MDA-MB-231 cells cultured in collagen ECM with concentration of 1.5 mg/mL (see \textit{SI appendix} S1). Consistent with previous reports, Y27632 reduces actomyosin contractility, promoting transitions from blebbing to mesenchymal phenotypes \cite{Janmey2005}. On the other hand, CN03 elevates myosin II activity, leading to retraction of filopodia to rounded cell shapes (Fig. \ref{fig6}A). We find that after the treatment, occurrence probabilities of morphological phenotypes are markedly distinct from the native (non-treated) cells (Fig. \ref{fig6}B). Perturbing the Rho signaling in both directions encourage the cells to sample their shape space more dynamically (Fig. \ref{fig6}C-D). As a result, both Y27632 and CN03 treatments allow the cells to more frequently visit otherwise hard to access states such as FP and AE. Invigorated by the perturbations, MDA-MB-231 cells exhibit reduced disparity in the phenotype occurrence probabilities compared with the native (non-treated) group. 

Tracking the phenotype evolution at single cell resolution further reveals the shape trajectories of MDA-MB-231 cells under perturbations (Fig. \ref{fig6}E). Despite fluctuations among individual cells, we find strong convergence of trajectories in cell shape space (Fig. \ref{fig6}E thick arrows). These converged paths are presumably the most biomechanically favorable. Consistent with our results in Fig. \ref{fig5}, AE state is heavily accessed as a gateway to FP state when cells are treated with Y27632. When treated with CN03, AE is at the fork of two bifurcating directions where one lead to LP, and the other lead to HB/SB states. These results suggest that AE state, which exhibits weak cell-ECM adhesions and F-actin rich protrusions \cite{Soldatl2006,Sahai2006pseudopodia}, mediate Rho-signaling controlled transitions between mesenchymal and amoeboidal motility.

Having analyzed the morphodynamics of MDA-MB-231 cells in 3D
matrices, we next examine the invasion potential of each morphological
phenotype. To this end, we model the migration of a cell as a
polarized persistent random walk \cite{Wirtz2014}. Because the finite
dwell time of each phenotype limits the length of trajectories, we
measure step size distribution as an alternative of autocorrelation
analysis in order to extract the diffusivity and persistence of cell
motility for each phenotype.

We consider one-hour segments of cell trajectories
divided into two half-hour sections. Assuming the average cell
velocity during the first section is along $\hat{x}$ direction, we
find the displacement $\delta\mathbf{d}$ of the cell in the second section well
described as 
\begin{eqnarray}
\delta\mathbf{d} &=& d_\parallel\hat{x} + d_\perp\hat{y} \nonumber\\
&=& W(v_\parallel\delta t,D_\parallel\delta t)\hat{x}
+ W(0,D_\perp\delta t)\hat{y},
\end{eqnarray}

where $\delta t$ = 0.5 hour, $W(\mu,\sigma^2)$ is a Wiener process of drift $\mu$ and
variance $\sigma^2$ (See also \textit{SI Appendix} S6). $D_\parallel$ and $D_\perp$ are effective
diffusion coefficients of the cell in two orthogonal directions: one
that is parallel to the previous step, and one that is perpendicular
to the previous step. $v_\parallel$ is the persistent velocity, which
quantifies the tendency of a cell to keep its migration direction.

\begin{figure}[h]
	\centering \includegraphics[width=0.99\columnwidth]{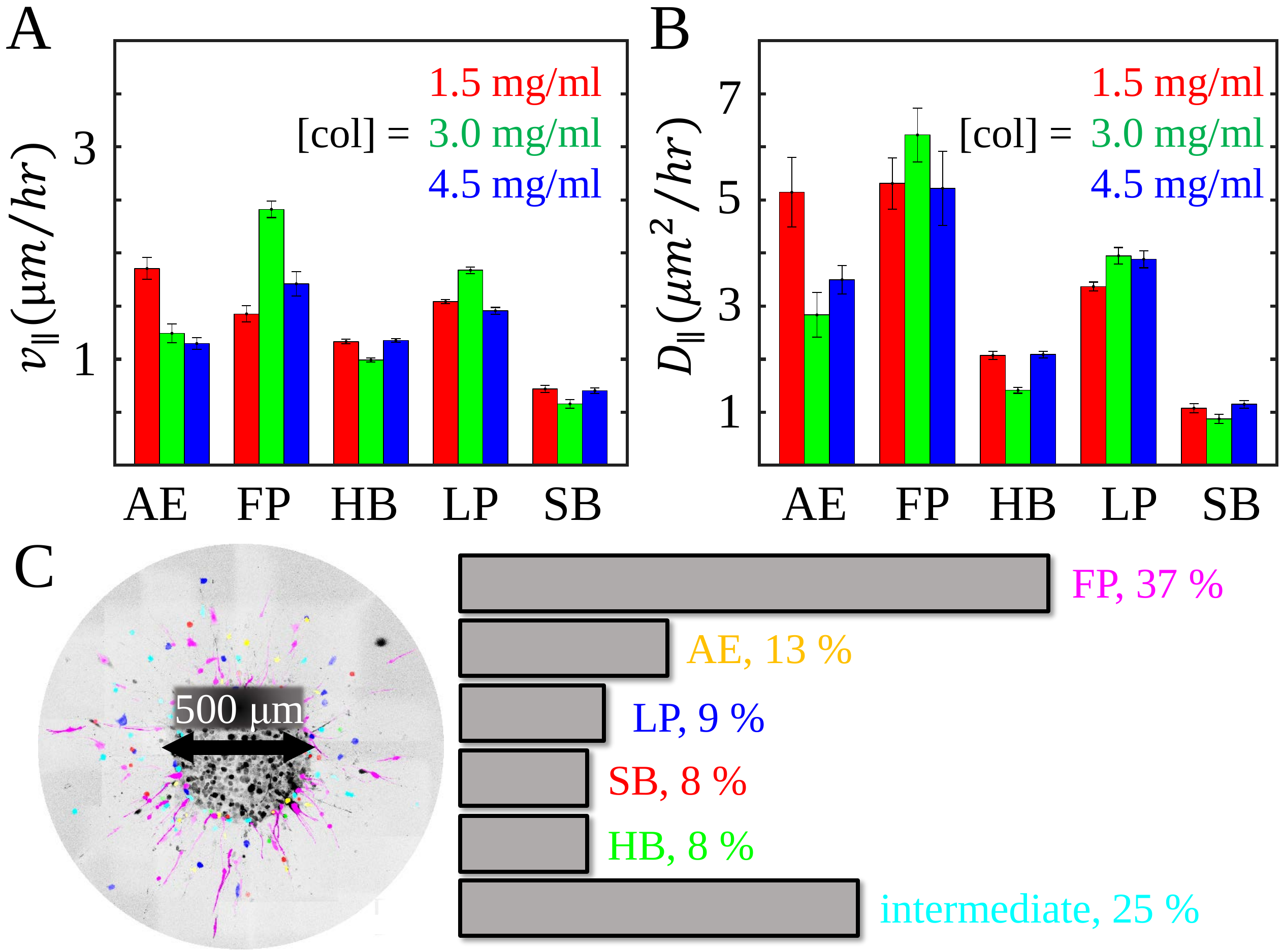}
	\caption{The morphological phenotype determines 3D invasive potential
		of MDA-MB-231 cells. (A-B) Persistence and diffusivity of each
		morphological phenotype in ECM of varying collagen
		concentrations. For each condition, $v_\parallel$ and $D_\parallel$
		are obtained by fitting the cumulative distribution function of
		samples with $N>$1000. The errorbars are 95\% confidence intervals
		of the fitting parameters. See also \textit{SI Appendix} S6 for
		more details. (C) Phenotype composition of cells disseminated from
		the original tumor organoid at day 5 after seeding the cells. Left: a confocal imaging snapshot of an organoid taken at day 5. Cells that have disseminated into the ECM are color-labeled by their phenotype classification. Right: phenotype composition of disseminating cells.}
	\label{fig7}
\end{figure}

We find overall filopodial cells have the highest rate of persist
migration (Fig. \ref{fig7}A-B and \textit{SI Appendix} S6),
while SB mode is the least motile. The phenotype-dependent motility is
also tuned by the ECM mechanics. In particular, $v_\parallel$ and
$D_\parallel$ are maximal at intermediate collagen concentration for
mesenchymal modes (3 mg/ml), but the trend is not present for
amoeboidal modes. Previously it was reported that MDA-MB-231 cell
invasion is most efficient at intermediate collagen concentration
\cite{Mierke2011}. Here we show that mesenchymal motility is mostly responsible
for such non-monotonic dependence.

Persistent motility predicts the invasion potential of cancer
cells. Because of the observed phenotype transitions
(Fig. \ref{fig5}), we expect that phenotype homeostasis is coupled
with invasion. In particular, we hypothesize a population selected for
higher invasion potential will have different phenotype occurrence
probability compared with the random migration assays in
Fig. \ref{fig4}.

To test the hypothesis, we study the invasion of a tumor diskoid into
ECM of 1.5 mg/ml type I collagen \cite{Alobaidi2017}. The diskoid of
500 $\mu$m in diameter consists of $\approx$ 1500 tightly packed
MDA-MB-231 cells. Over the course of 5 days, cells disseminate into the
ECM as deep as 400 $\mu$m from the original tumor. Fig. \ref{fig7}C
shows a snapshot taken on the fifth day of
the invading diskoid (See also \textit{SI Appendix} S7).

We examine the morphological phenotype of all cells that have
disseminated from the original tumor. Consistent with our hypothesis,
this more invasive population favors AE and FP modes, which have the
highest persistent motility at [col]=1.5 mg/ml as shown in
Fig. \ref{fig7}A-B. On the other hand, the slow moving SB modes only
account for 8\% of the population disseminated from the original tumor
organoid. The shift of phenotype homeostasis is especially striking
when comparing Fig. 4B and Fig. 6C, where the most and least
probable phenotypes are opposite in these two situations.

\section{Discussion}
\label{sec:discuss}
In this paper, we report the morphodynamics of MDA-MB-231 cells in
type I collagen ECM as a model system of metastatic cancer cells
migrating in 3D tissue. MDA-MB-231 cells rapidly sample their possible
geometry, therefore exhibiting simultaneous random walks in both real
space and shape space (Fig. \ref{fig1}). When ECM mechanics are tuned
by varying collagen concentration, we find dramatic changes in the
cellular morphodynamics, but not in the ensemble distributions of cell
shapes (Fig. \ref{fig2} and Fig. \ref{fig3}). On the other hand, when
the cells migrate on the surfaces of collagen ECM, their
shape distributions do shift as collagen concentration increases. The
contrast is particularly striking noticing that almost the same set of molecular machineries are involved in both 2D and 3D cell migration, shape regulation and mechanosensing \cite{Yamada2016adhesion,Ballestrem2017}.

The biological significance of the morphodynamics is further
demonstrated by classifying cell shapes into morphological phenotypes
corresponding to different migration programs. While the occurrence
and dwell time of blebbing phenotypes increases with collagen
concentration, mesenchymal shapes become less stable and probable
(Fig. \ref{fig4} and Fig. \ref{fig5}). Interestingly, accompanying
these two opposite trends is the increased transition rates among
morphological phenotypes that are elevated at higher collagen
concentration (Fig. \ref{fig5}).

Our results also shed light to the control mechanism of cell
motility phenotypes. In particular, we show that perturbing the Rho-signaling dramatically modulates the cell morphodynamics (Fig. \ref{fig6}). It has
been shown previously that Rac and Rho signaling regulates the shift
between lamellipodial and blebbing motility
\cite{Marshall2008rac,Marshall2010plasticity}. Complement to these results, our analysis further reveal that instead of stabilizing a particular mode of motility, perturbation of Rho signaling invigorates the morphodynamics of cells. As a result, cells exhibit more dynamic shape fluctuations and more aggressively explore the different morphological phenotypes. We also find that AE state mediate the Rho-signaling controlled transitions between different phenotypes, further illuminating the underlying phenotype landscape that controls cancer cell 3D motility.

In light of the rapid phenotype transitions exhibited by individual
cells, 3D cancer cell motility may be considered as a hidden Markov
process where each phenotype is associated with characteristic
diffusivity and persistence. In this perspective morphodynamics
facilitates cancer invasion because phenotype transitions allow cancer
cells to search for and commit to a more invasive phenotype (Fig. \ref{fig7}), \cite{Lander2011}.

In summary, we demonstrate the morphodynamics of 3D migrating cancer
cells as a powerful tool to inspect the internal state and
microenvironment of the cells. In order to further exploit the
information provided by the cell shape fluctuations, future research
is needed to decode morphodynamics as a rich body language of cells,
and to control morphodynamics as a route of mechanical programming of
cell phenotype.

\section{Materials and methods}
\label{sec:methods}
See  \textit{SI Appendix} S1 for details of 3D cell culture, microscopy, and pharmacological treatments.

\begin{acknowledgments}
	We thank Prof. Michelle Digman and Prof. Steve Press{\'e} for helpful
	discussions. The funding for this research results from a Scialog
	Program sponsored jointly by Research Corporation for Science
	Advancement and the Gordon and Betty Moore Foundation through a grant
	to Oregon State University by the Gordon and Betty Moore Foundation
\end{acknowledgments}

\bibliography{morphodynamics_arxiv}

\end{document}